\documentclass[twocolumn,aps,prl,preprintnumbers,superscriptaddress,nofootinbib,10pt]{revtex4-1}
\usepackage{mathrsfs}
\usepackage{amsfonts}
\usepackage{amsmath}
\usepackage{amssymb}
\usepackage{array}
\usepackage{verbatim}
\usepackage{epsfig}
\usepackage{graphicx}
\usepackage{amsbsy}
\usepackage{bm}
\usepackage[dvipsnames]{xcolor}  
\usepackage{graphicx} 
\usepackage[font=small,labelfont=bf]{caption} 
\usepackage[subrefformat=parens]{subcaption}

\usepackage{import}
\input{epsf}
\usepackage{psfrag}
\usepackage{xcolor}
\usepackage{slashed}
\usepackage{hyperref}
\usepackage{ulem}
\usepackage{MnSymbol}

\begin{document}

\begin{flushright}

\end{flushright}

\title{Entanglement Negativity of Spin-Orbit Correlations in a general Qubit-Qudit Setup}

\author{Sanskriti Agrawal}
\affiliation{Department of Physics, Aligarh Muslim University, Aligarh - $202001$, India.}

\author{Raktim Abir}
\affiliation{Department of Physics, Aligarh Muslim University, Aligarh - $202001$, India.}
\email{raktim.ph@amu.ac.in}

\begin{abstract}

We present the complete eigenvalue spectrum of the partially transposed density matrix for a pure bipartite quantum state acting on a generic $2 \otimes n$ Hilbert space. The spectrum contains four non-zero eigenvalues, as, 
\begin{eqnarray}
\lambda_{1,2}=\pm \sqrt{A}, ~~~ \lambda_{3,4}= \frac{1}{2}(1\pm\sqrt{1-4 A}),  \nonumber 
\end{eqnarray}
where $A$ is the determinant of the reduced density matrix (traced over the larger subspace). As $0 \leqslant A \leqslant1/4$, only one is negative among the four non-trivial eigenvalues. 
Within this qubit-qudit framework, we further studied the negativity as a measure of entanglement for the case of spin-orbit correlation of partons inside a proton. The entanglement negativity for spin-orbit correlations is found to be related to the gluon helicity PDF and the Hermitian angle of the associated Hilbert space for linearly polarized protons.

\end{abstract}

\maketitle
{\color{Red} $\bullet$} {\color{NavyBlue} {\it Introduction}:~}
The classical {\it bits} and their quantum counterpart {\it qubits} are two-level systems with basis states $e.g.$ $\{0,1\}$ and $\{ |0\rangle, |1\rangle\}$ respectively.
The complexity of nature however almost always makes the physical system multi-level. This requires a higher-dimensional Hilbert space to study the quantum evolutions and quantum correlations of those physical systems. One such physical situation that requires a multi-dimensional Hilbert space is the orbital angular momentum (OAM)  states, which require $2l+1$ states for each value of angular momentum $l$. A simple qubit can't accommodate them. While a qubit is for the binary digits, a qudit is a quantum manifestation of the $n$-ary digits.  The state of the qudit can be described by a set of orthonormal basis states, $e.g.$, in the computational basis as $\{|0\rangle, |1\rangle, ..., |n-1\rangle\}$. When the angular momentum entangles with spin degrees of freedom, one requires a qubit for the spin and a qudit for the orbital angular momentum to set a general framework for studying the quantum dynamics of the composite Spin-OAM system.

Spin-orbit entanglement refers to bipartite entanglement between two internal degrees of freedom - spin and orbital angular momentum of a particle. The particle can either be spin-1 $e.g.$ photons or gluons, or, spin-$1/2$ $e.g.$ quarks and leptons.
Being a non-local quantum correlation, the entanglement of a quantum state remains invariant under local unitary transformations. Therefore, a wide range of operations that can be attributed to unitary transformations $e.g.$, free space propagation of the particles, will not generally affect the entanglement between the internal degrees of freedom of that particle. 
Nevertheless, it is possible to quantify the entanglement by computing the entanglement measures.

In this article, we presented the eigenvalue spectrum of the partial transposed density matrix of a general $2 \otimes n$ qubit-qudit system. While all our results are for the general $2 \otimes n$ system, our presentation revolved around $2 \otimes (2l+1)$ as this sub-class resembles the spin-orbit quantum correlations of sub-atomic particles.
\\

{\color{Red} $\bullet$} {\color{NavyBlue} {\it Eigenvalue spectrum of partial transposed $2\otimes n$ bipartite system}:~} 
We begin with a generic pure state ${\bm \phi} \in {\cal H}_s^2 \otimes {\cal H}_l^{2l+1}$. Given the orthonormal basis 
$\{|+\rangle , |-\rangle\}$ and $\{|l\rangle, ..., |-l\rangle\}$ for spin and orbital angular momentum, the state  ${\bm \phi}$ can be written as \cite{Hatta:2024lbw}, 
\begin{eqnarray} 
|{\bm \phi} \rangle 
&&= \sum_{s = +,-} ~\sum_{l = l,\dots ,-l} |s\rangle \otimes |l\rangle \equiv  \sum_{s,l}  c_{s,l}|s, l\rangle , \nonumber \\
&&= \{ c_{+,l}|+,l\rangle  ... + c_{+,0}|+,0\rangle ... + c_{+,-l}|+,-l\rangle  \} \nonumber\\
&&~~~~+  \{ c_{-,l}|-,l\rangle  ... + c_{-,0}|-,0\rangle ... + c_{-,-l}|-,-l\rangle  \}.  \label{quin2}
\end{eqnarray}
Here, $|s \rangle$ represents the spin eigenstates of the particles and $|l\rangle$ represents the OAM eigenstates with eigenvalues ranging from $l$ to $-l$. The coefficients $c_{s,l}\in {\mathbb C}$ and normalization of the state implies,
\begin{eqnarray} 
\langle {\bm \phi}|{\bm \phi} \rangle = \sum_{l} |c_{+,l}|^2 + |c_{-,l}|^2 = 1. \label{quin9}
\end{eqnarray} 
%
%
From the state $|\bm\phi \rangle$, one may define the associated density matrix as, $\rho = |{\bm \phi} \rangle  \langle {\bm \phi}|$. Density matrices defined this way are pure states. Pure states always have one eigenvalue equal to unity; all other eigenvalues are zero.
In case of the spin-orbit correlations, $\rho$ is a $(4l+2) \times (4l+2)$ dimensional Hermitian matrix with the elements as follows, 
\begin{eqnarray}
\rho = 
\begin{pmatrix}
& c_{+,l}^{}~c_{+,l}^{*} & & c_{+,l}^{}~c_{+,(l-1)}^{*} &  ... &  c_{+,l}^{}~c_{-,-l}^{*} &  \\ \\ 
& c_{+,(l-1)}^{}~c_{+,l}^{*} && c_{+,(l-1)}^{}~c_{+,(l-1)}^{*} &  ... &  ... &  \\  \\
& \vdots && \vdots  & \ddots & \vdots \\ \\
& c_{-,-l}^{}~c_{+,l}^{*} && ... &  ... &  c_{-,-l}^{}~c_{-,-l}^{*} & \\   \label{quin1}
\end{pmatrix}.
\end{eqnarray} 
When traced over the $l$-subspace, $\rho$ becomes $2\times2$ {\it{reduced density matrix}} $\rho_l$,
\begin{eqnarray} 
\rho_l={\rm Tr}_{l}|{\bm \phi} \rangle  \langle {\bm \phi}| = 
\begin{pmatrix}
& \sum_{l} |c_{+,l}^{}|^2 & & \sum_{l} c_{+,l}^{} c_{-,l}^{*}  & \\ \\
& \sum_{l} c_{-,l}^{} c_{+,l}^{*} & & \sum_{l} |c_{-,l}^{}|^2 & \\
\end{pmatrix}. \label{quin3}
\end{eqnarray}
For any density matrix or quantum state $\rho$, it is a fact that $\rho$ and its transpose $\rho^T$ share the identical spectrum of eigenvalues. However, if $\rho \in  {\cal H}_m \otimes {\cal H}_n$ then the eigenvalue spectrum of the partial transpose of the density matrix, ($I_m \otimes T_n$)($\rho$), will generally be quite different from that of $\rho$. More specifically, in the case of a qubit-qudit system, the partial transpose of the general state $\rho$ acting on the Hilbert space ${\cal H}_s\otimes {\cal{H}}_{2l+1}$ can be written as, 
%
\begin{eqnarray}
    \rho^{T_l}&=& \left(I_2\otimes T_{2l+1}\right)(\rho), 
\end{eqnarray}
\noindent
where the transpose operation is made only in the $l$-subspace. If all eigenvalues of $\rho^{T_l}$ are non-negative, then $\rho$ is likely to be a separable state or atmost a bound entangled state. Otherwise, if there are one or more negative eigenvalue(s) present, then it is non-positive partial transposition (NPT). The NPT states are necessarily entangled through {\it Peres-Horodecki} criterion \cite{Peres:1996dw,Horodecki:1996nc}. To quantify this entanglement, one of the most useful measures is {\it entanglement negativity} introduced by Vidal and Werner \cite{Vidal:2002zz}. The {\it negativity} is defined as the absolute sum of the negative eigenvalues of the partial transpose of $\rho$ ($\rho^{T_l}$). So, negative eigenvalues of $\rho^{T_l}$ not only confirm the existence of entanglement, but also quantify the amount of entanglement in $\rho$. Thus, it is important and interesting to study the negativity of the density matrix $\rho$ by exploring the negative eigenvalues of $\rho^{T_l}$. The partially transposed density matrix $\rho^{T_l}$ has $4l+2$ eigenvalues. We have found that partially transposed density matrix of a general $2 \otimes n$ (qubit-qudit system) has only four non-zero eigenvalues, rest of the eigenvalues are identically zero, as, 
\begin{eqnarray}
\lambda_1&=&-\sqrt{A},\nonumber\\
\lambda_2&=&\sqrt{A},\nonumber\\
\lambda_3&=&\frac{1}{2}(1-\sqrt{1-4 A}),\nonumber\\
\lambda_4&=&\frac{1}{2}(1+\sqrt{1-4 A}),\nonumber\\
\lambda_{5}\dots\lambda_{4l+2}&=&0,\label{quin15}
\end{eqnarray}
\noindent where, $A = \text{det}(\rho_l)$, with $\rho_l$ being the reduced density matrix. 
\begin{figure}
    \centering
    \includegraphics[width=0.8\linewidth]{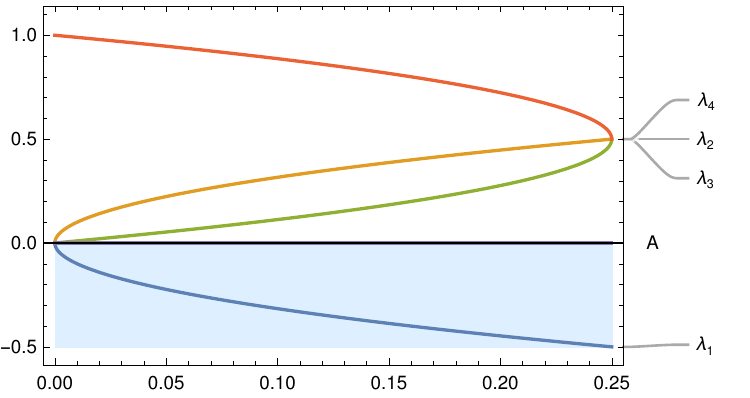}
    \caption{The four non-zero eigenvalues of partially transposed $\rho^{T_l}$ plotted as function of $A$. While separable states belong to $A=0$, the maximally entangled states is at $A=1/4$. When the state is maximally entangled, the three positive eigenvalues $\lambda_2, \lambda_3, \lambda_4$ become degenerate at value $1/2$ and the negative eigenvalue $\lambda_1$ saturates to $-1/2$.}
    \label{fig1}
\end{figure}
While a general proof is elusive, we have presented the details for the qubit-qubit $(2\otimes 2)$, the qubit-qutrit $(2 \otimes 3)$, and an outline for general $(2\otimes n)$ in the additional notes. One may further confirm the results in the computational platform, $e.g.$ {\it Mathematica} for higher values of $n$.  
Both the density matrix and it's partial transpose are Hermitian matrices having only real eigenvalues. This will ensure, $A\geqslant0$ else $\lambda_1$ and $\lambda_2$ will become complex and $A\leqslant{1/4}$, else $\lambda_3$ and $\lambda_4$ will become complex. As, $A$ is bounded between $0$ and $1/4$, the spectrum contains only one negative eigenvalue $\lambda_1=-\sqrt{A}$ as in Fig. \ref{fig1}. The characteristic equation of $\rho^{T_l}$ is $(\lambda^2-A)(\lambda^2-\lambda+A)\lambda^{4l-2}=0$ plotted in Fig \ref{fig2}. It is also interesting to see that, 
\begin{eqnarray}
\sum_{i=1}^{4l+2} \lambda_i = 1,  ~~~~~
\sum_{i=1}^{4l+2} \lambda_i^2 = 1, 
\end{eqnarray}
which refers to the fact that ${\rm Tr}\left(\rho^{T_l}\right) =1$ and ${\rm Tr}\left(\rho^{T_l}\right)^2=1$.  
The factor $A$, 
\begin{eqnarray}
    A = \text{det}(\rho_l)= \sum_l |c^{}_{+,l}|^2~\sum_l |c^{}_{-,l}|^2 -\left(\sum_l c^{}_{+,l} c^*_{-,l} \right)^2 \geqslant 0, \label{quin16}
\end{eqnarray}
is understood to be positive as $c_{\pm,l}$'s are complex numbers, and the inequality follows from the {\it Cauchy-Schwarz} inequality. \\

{\color{Red} $\bullet$} {\color{NavyBlue} {\it Entanglement Negativity}:~} 
Cauchy-Schwarz inequality is essentially an upper bound on the inner product between two vectors in an inner product space in terms of the product of the vector norms. Both $\{c_{+,l}\}$ and $\{c_{-,l}\}$ can be taken as vectors in the $2l+1$ complex space as, ${\bm c_+,\bm c_-}\in{\mathbb C}^{2l+1}$ with ${\bm c_+}=\left(c_{+,l},...,c_{+,0},...,c_{+,-l}\right)$ and ${\bm c_-}=\left(c_{-,l},...,c_{-,0},...,c_{-,-l}\right)$. The canonical complex inner product on the vector space ${\mathbb C}^{2l+1}$ is defined by $\langle\bm c_+,\bm c_-\rangle= c^{}_{+,l}{ c^*_{-,l}}+...+c^{}_{+,0}{ c^*_{-,0}}+...+c^{}_{-,-l}{ c^*_{-,-l}}$.
\begin{figure}
    \centering
    \includegraphics[width=0.85\linewidth]{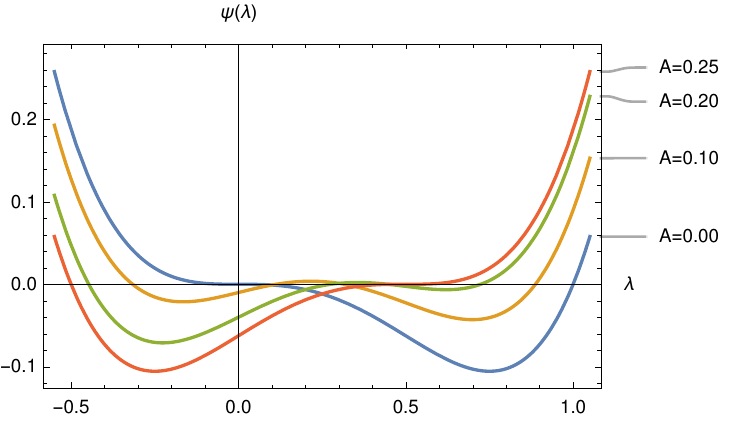}
    \caption{The characteristics polynomial $\psi(\lambda)=\left(\lambda^2-A\right)\left(\lambda^2-\lambda+A\right)$ plotted as the function of $\lambda$ for the full range of $A$. The zero's of $\psi(\lambda)$ are the eigenvalues of partial transposed $\rho$ $(\rho^{T_l})$. One can see, other than $A=0$, which may correspond to separable states, all other cases have a single negative eigenvalue, confirming the entanglement of the state $\rho$.}
    \label{fig2}
\end{figure}
With the norms and inner product now defined, the reduced density matrix $\rho_l$ can be expressed as,
\begin{eqnarray} 
\rho_l = 
\begin{pmatrix}
&  ||\bm c_+||^2 & & \langle \bm c_+,\bm c_-\rangle  & \\ \\
& \langle \bm c_-,\bm c_+\rangle & & ||\bm c_-||^2& \\
\end{pmatrix}. \label{quin3}
\end{eqnarray}
As the eigenvalue spectrum of $\rho^{T_l}$ contains only one negative eigenvalue, the negativity of the spin-orbit correlation is 
\begin{eqnarray} 
    {\cal N} (\rho )
    &=& \sqrt{||\bm c_+||^2 ||\bm c_-||^2-|\langle \bm c_+,\bm c_-\rangle|^2} = ||{\bm c}_{+} \wedge {\bm c}_{-} ||.
\end{eqnarray}
Negativity therefore, is the norm of the wedge product of the two vectors ${\bm c}_+$ and ${\bm c}_-$. It vanishes only when ${\bm c}_+$ and ${\bm c}_-$ are `collinear' $i.e.$ ${\bm c}_+ \propto {\bm c}_-.$  
Further, we can introduce the angle between complex vectors $\bm c_+$ and $\bm c_-$ as,
\begin{eqnarray}
    \cos\theta_H =  \frac{|\langle\bm c_+,\bm c_-\rangle|}{||\bm c_+||~||\bm c_-||}.
\end{eqnarray}
The angle $\theta_H$ is known as the Hermitian angle between ${\bm c}_+$ and ${\bm c}_-$ in the complex vector space. Thus, the negativity can be written as,
\begin{eqnarray}
     {\cal N} (\rho )&=& ||\bm c_+||~ ||\bm c_-|| \sin\theta_H.
\end{eqnarray}
Negativity being square-root of determinant of the reduced density matrix is also consistent with the Schmidt decomposition theorem, as for any pure bipartite state in ${\cal H}_A^{n_1}\otimes {\cal H}_B^{n_2}$ ($n_1 <n_2$), the negativity can be written as, $\sum_{i<j}\sqrt{\alpha_i \alpha_j},$
where, $\sqrt{\alpha_i}$ and $\sqrt{\alpha_j}$ are Schmidt coefficients, with $\alpha_i$ and ${\alpha_j}$ are the eigenvalues of the reduced density matrix $\rho_A$ \cite{Lee_2003}. 
Finally, we use Lagrange's identity to get, 
\begin{eqnarray}
     {\cal N} (\rho ) &=& \sum_l |c^{}_{+,l}|^2~\sum_l |c^{}_{-,l}|^2 -\left|\sum_l c^{}_{+,l} c^*_{-,l} \right|^2  \nonumber \\
     &=& \sum_{-l \leqslant  i<j \leqslant  l} \left|c^{}_{+,i} c^{*}_{-,j} - c^{}_{+,j}  c^{*}_{-,i}\right|^2 \equiv \sum_{-l \leqslant  i<j \leqslant  l} \tau_{i,j} , \label{quin16}
\end{eqnarray}
There are $l(2l-1)$ numbers of $\tau_{ij}$. Each of them is either zero or positive. Therefore, to confirm or certify the entanglement, it's sufficient to show that at least one $\tau_{ij}$ is non-zero for some $i,j \in \{+l,...,-l\}$ and $i\neq j$, $i.e.$, 
\begin{eqnarray}
c^{}_{+,i} c^{*}_{-,j} - c^{}_{+,j} c^{*}_{-,i} \neq 0.
\end{eqnarray}
As $\langle s, l | \phi \rangle = c_{s, l}$, we define the operator $\tilde{\cal E}_{ij}$ for any pair of $i,j \in \{l,...,-l\}$ with $i\neq j$ as, 
\begin{eqnarray}
\tilde{\cal E}_{ij} = |-,j\rangle \langle +, i | - |-,i\rangle \langle +, j|, 
\end{eqnarray}
For some choice of $i,j \in \{l,...,-l\}$ with $i\neq j$, if,
\begin{eqnarray}
  \langle  \phi | \tilde{\cal E}_{ij} | \phi \rangle \neq 0, 
\end{eqnarray}
that will be sufficient to certify the entanglement as the negativity will be non-zero. 
\noindent
Negativity is an entanglement monotone. Therefore, it not only quantifies the amount of entanglement of a quantum state, it is also a non-negative function whose value does not increase under any local operations and classical communications.
It is interesting to see that entanglement entropy of $\rho$, which is essentially the von-Nuemann entropy of the reduced density matrix $\rho_l$, 
\begin{eqnarray}\nonumber
    { \cal E}(\rho)&\equiv& - \sum_{i=+,-} \lambda_i\log \lambda_i,\\
    &=& -\frac{1}{2}\log{A}-\frac{1}{2}~\sqrt{1-4A} ~\log\left[\frac{1+\sqrt{1-4A}}{1-\sqrt{1-4A}}\right],
\end{eqnarray}
is function of the single variable $A=\det(\rho_l)$. The determinant of the reduced density matrix is therefore acts as an exclusive single parameter to quantify the entanglement. At $A\rightarrow0$, entanglement entropy ${\cal E}(\rho)$ is regular and vanishes identically. When the state is maximally entangled ($A=1/4$), ${\cal E}(\rho)$ reaches its maximum value at $\ln2$.
Another important entanglement measure of the pure state $\rho$ is the {\it concurrence}, $\cal C$ \cite{Hill:1997pfa,Wootters:1997id},
\begin{eqnarray}
    {\cal C}\left(\rho\right)\equiv \sqrt{2\left(1-{\text Tr}\left(\rho^2_l\right)\right)}= 2 \det (\rho_l). \label{quin20}
    \end{eqnarray}
  The reduced density matrix being a $2\times 2$ matrix, the last equality comes from the determinant-trace relation of square matrix, ${\text {det }M}=\frac{1}{2}\left(({\text {tr }M)^2 -{\text {tr }M^2}}\right)$. Therefore, for the qubit-qudit system, negativity and concurrence are related as,  
  \begin{eqnarray}
      {\cal C}(\rho)=2{\cal N}^2(\rho).
  \end{eqnarray}
  This indeed confirms that the partial transpose acts as the direct link between negativity and concurrence \cite{Eltschka_2015}.\\

{\color{Red} $\bullet$} {\color{NavyBlue} {\it Maximally entangled state}:~}The spin-orbit composite system contained in the density matrix $\rho$ is in pure state. However, the spin sub-system (traced over the OAM space) contained in $\rho_l$ is a mixed state. This can be confirmed from the $\it purity$ of reduced density matrix $\rho_l$,
\begin{eqnarray}\nonumber
    \gamma(\rho_l)&\equiv&{\text Tr}(\rho^2_l)\\
    &=&||\bm c_+||^4+||\bm c_-||^4+2|\langle\bm c_+,\bm c_-\rangle|^2 <{1}.
\end{eqnarray}
While, pure state has $\gamma=1$, mixed states have $\gamma<1$. It is interesting to see that the negativity of the spin-orbit composite system and the purity of the spin sub-system are related as,
\begin{eqnarray}
    \gamma(\rho_l)=1-2{\cal N}^2(\rho).
\end{eqnarray}
We see from the above expression that the more entangled the spin-orbit composite state is ($i.e.$ more negativity), the less pure is the spin sub-system ($i.e.$ more mixed).
When a pure bipartite state is maximally entangled, its reduced density matrix is maximally mixed. A mixed state is a convex combination of pure states. Hence, a mixed state can be written as, $\rho =\sum_i p_i |\phi_i\rangle \langle\phi_i|$, where, $p_i \geqslant 0$, $\sum_i p_i =1$. These $p_i$'s can be thought of probability distribution of individual pure states. If we assume that $\phi_i$'s form an orthonormal basis, then a maximally mixed state is a state where $p_i$'s are uniformly distributed, $i.e.$, $p_i=1/n$, where $n$ is the dimension of the state. Therefore, the maximally mixed states are mixed states where a mixture of all states offers the same probability.
This leads to the fact that the associated reduced density matrix is proportional to the identity matrix, $\rho_l=(1/d) I_{d\times d}$.
Therefore, the spin-orbit correlation is maximally entangled if the associated density matrix $\rho_l$ is maximally mixed and has the form,
$\rho_l=(1/2) I_{2\times2}$.
This requires the following two conditions to be fulfilled in Eq.\eqref{quin3},
\begin{itemize}
    \item $||\bm c_+||^2=||\bm c_-||^2 =1/2$, $i.e.$, $\bm c_+$ and $\bm c_-$ are of identical norm and,
    \item $\langle \bm c_+,\bm c_-\rangle=\langle \bm c_-,\bm c_+\rangle=0$, $i.e.$, $\bm c_+$ and $\bm c_-$ are orthogonal. 
\end{itemize}
 The above two conditions will ensure the maximal entanglement of the spin-orbit system, having entanglement entropy and entanglement negativity saturating to their maximum values at $\ln 2$ and $1/2$ respectively.\\

{\color{Red} $\bullet$} {\color{NavyBlue} {\it Polarized and Unpolarized Hadrons}:~}
If the hadron is taken to be unpolarized, two conditions follow \cite{Hatta:2024lbw},
\begin{eqnarray}
    \langle \bm\phi|s^z|\bm\phi\rangle=0, ~~~
    \langle\bm\phi|l^z|\bm\phi\rangle=0. \label{quin7}
\end{eqnarray}
Even in an unpolarized hadron, certain correlation between spin and orbital angular momentum may present as, $\langle\bm\phi|s^zl^z|\bm\phi\rangle\propto C_{q,g}(x)$, where, $C_{q,g}(x)$ are (quarks and gluons) spin-orbit correlations \cite{Bhattacharya:2024sno}.
%
%
As consequence of Eq.\eqref{quin7}, one can write,
\begin{eqnarray}
    \sum_l\left(| c_{+,l}|^2-|c_{-,l}|^2\right)=0,~~~
    \sum_ll\left(|c_{+,l}|^2+|c_{-,l}|^2\right)=0. \label{quin10}
\end{eqnarray}
For unpolarized hadrons, it has been argued that for individual partons, the spin and orbital angular momentum are maximally entangled, making them belong to the Bell states \cite{Bhattacharya:2024sno,Hatta:2024lbw}. From Eq.\eqref{quin9} and \eqref{quin10}, we can further write,
\begin{eqnarray}
    \sum_l | c_{+,l}|^2=\frac{1}{2},~~~ \sum_l |c_{-,l}|^2=\frac{1}{2}, \label{quin11}
\end{eqnarray}
$i.e.$, $||\bm c_+||=||\bm c_-||=1/\sqrt{2}$.
Therefore, partons in an unpolarized proton, the first of the two conditions, as to make the state maximally entangled, is automatically fulfilled. What remaines is to have $\bm c_+$ and $\bm c_-$ orthogonal, $i.e.$, $\langle\bm c_+,\bm c_-\rangle=0$.
Moving to the case of polarized hadrons where $||\bm c_+||\neq ||\bm c_-||$, unlike the unpolarized case $\langle\bm\phi|s^z|\bm\phi\rangle$ is not identically zero, rather is proportional to the gluon helicity PDF,
\begin{eqnarray}
    \sum_l\left(|c_{+,l}|^2-|c_{-,l}|^2\right)=\frac{\Delta g(x)}{g(x)},\label{quin4}
\end{eqnarray}
where, $\Delta g(x)$ is gluon helicity PDF function, $g(x)$ is gluon PDF.
%
As a consequence the reduced density matrix will no longer be maximally mixed.  Therefore, the entanglement negativity of spin-orbit correlation for the case of polarized hadrons is, 
\begin{eqnarray}\nonumber
    {\cal N}(\rho)&=&||\bm c_+||~||\bm c_-||\sin\theta_H,\\ 
    &=&\frac{1}{2}\left(1-\frac{\left(\Delta g(x)\right)^2}{g^2(x)}\right)^{1/2}\sin\theta_H.
\end{eqnarray} 
This relates the entanglement negativity and the hermitian angle with the parton distribution functions for polarized hadrons.  \\

{\color{Red} $\bullet$} {\color{NavyBlue} {\it Conclusion}:~} The entries of the reduced density matrix $\rho_l$ are either norms or inner products. As norms and inner products are invariant under unitary transformations, $\langle U({\bm c_+}),U({\bm c_-}) \rangle  =\langle\bm c_+,\bm c_-\rangle$, any unitary transformation over $\bm c_+$ and $\bm c_-$ will keep $\rho_l$ to remain invariant, so as the entanglement entropy or negativity. Unitary operations are automorphisms of the inner product spaces, $i.e.$, they preserve the structure and invariants of the space on which they act. Going in the small-$x$ limit is essentially the high energy limit of the states. The boost in the $z$-direction, $e^{i\omega k^3}|\bm\phi\rangle$  is a unitary transformation. While this transformation may change the individual values of $ c_+$'s and $ c_-$'s, as well as the matrix $\rho$, however, being unitary tranformation they keep the norms and inner product invariant. Therefore, the reduced density matrix, entanglement entropy, and negativity will retain their initial values and do not depend on the Bjorken-$x$ as the state evolves at high energy. If the state is maximally entangled in the limit $x \rightarrow 0$, then the states with $0 < x < 1$ evolve from the maximally entangled Bell state via successive local unitary transformations are also maximally entangled \cite{Hatta:2024lbw}. Such maximal entanglement in the proton structure exhibits a relation between the jet fragmentation function and the entropy of hadrons produced in the scattering events \cite{Datta:2024hpn}. The CHSH inequality in the context of fragmentation of a single parton to hadron pairs is also used to understand the quantum nature of the hadronization process \cite{vonKuk:2025kbv}. 

Recent years have seen a surge in interest and activities in the interface of quantum information science and quantum chromodynamics \cite{Kutak:2011rb, Peschanski:2012cw,Kovner:2015hga, Kharzeev:2017qzs, Armesto:2019mna, Ramos:2020kaj, Beane:2018oxh, Liu:2022hto}. Most of the entanglement studies, in this interface, so far revolved mainly around the entropy, especially the entanglement entropy, except for a few recent studies \cite{Dumitru:2025bib}. The presence of negative eigenvalues in the spectrum of the partially transposed density matrix also confirms the entanglement of the quantum state. The negative eigenvalue(s) quantify the amount of entanglement through novel entanglement measures, $e.g.$, negativity.
In this paper, we have presented a complete characterization of the eigenvalue spectrum of the partially transpose density matrix for a $2\times (2n+1)$ qubit-qudit system \cite{Gerjuoy_2003}. As a measure of entanglement in the context of spin-orbit entanglement of partons, we have studied the negativity for the first time.  Our study is completely general and does not restrict the eigenstates of angular momentum. We found that the spectrum contains only one negative eigenvalue, which coincides with the square root of the determinant of the reduced density matrix. A single negative eigenvalue for $\rho^{T_l}$ is often interpreted as the amount of white noise required to make the state separable. It will be interesting to see how the negativity, containing the Hermitian angle, parameterizes the entanglement in the polarized events at colliders $e.g.$, in the upcoming Electron-Ion Collider. 

\bibliography{ref.bib}

\begin{widetext}

{\color{Red} $\bullet$} {\color{NavyBlue} {\it Appendix}:~}

In this note, we present the details of getting the characteristic equations for the partially transposed density matrix for the systems $2\otimes2$ and $2\otimes3$ and 
an outline to get it for the general $2\otimes n$ systems. \\

{\color{Red} $\bullet$} {\color{NavyBlue} {\it $2\otimes2$ system}:~}
Lets take the basis for one subsystem as $\{+,-\}$ and other as $\{+1,-1\}$. 
The density matrix for the $2\otimes2$ composite system can be written as,
\begin{eqnarray}
\rho = 
\begin{pmatrix}
& c_{+,1}^{}~c_{+,1}^{*} & & c_{+,1}^{}~c_{+,-1}^{*} & c_{+,1}^{}~c_{-,1}^{*}  &  c_{+,1}^{}~c_{-,-1}^{*} &  \\ \\ 
& c_{+,-1}^{}~c_{+,1}^{*} && c_{+,-1}^{}~c_{+,-1}^{*} & c_{+,-1}^{}~c_{-,1}^{*} &  c_{+,-1}^{}~c_{-,-1}^{*} &  \\  \\
& c_{-,1}^{}~c_{+,1}^{*} && c_{-,1}^{}~c_{+,-1}^{*}  & c_{-,1}^{}~c_{-,1}^{*} & c_{-,1}^{}~c_{-,-1}^{*} \\ \\
& c_{-,-1}^{}~c_{+,1}^{*} && c_{-,-1}^{}~c_{+,-1}^{*} &  c_{-,-1}^{}~c_{-,1}^{*} &  c_{-,-1}^{}~c_{-,-1}^{*} & \\   \label{quin1}
\end{pmatrix}.
\end{eqnarray} 
The partial transpose of the above density matrix will then be,
\begin{eqnarray}
\rho^{T_l} &=&
\begin{pmatrix}
& c_{+,1}^{}~c_{+,1}^{*} & & c_{+,1}^{}~c_{+,-1}^{*} & c_{+,1}^{}~c_{-,1}^{*}  &  c_{+,1}^{}~c_{-,-1}^{*} &  \\ \\ 
& c_{+,-1}^{}~c_{+,1}^{*} && c_{+,-1}^{}~c_{+,-1}^{*} & c_{+,-1}^{}~c_{-,1}^{*} &  c_{+,-1}^{}~c_{-,-1}^{*} &  \\  \\
& c_{-,1}^{}~c_{+,1}^{*} && c_{-,1}^{}~c_{+,-1}^{*}  & c_{-,1}^{}~c_{-,1}^{*} & c_{-,1}^{}~c_{-,-1}^{*} \\ \\
& c_{-,-1}^{}~c_{+,1}^{*} && c_{-,-1}^{}~c_{+,-1}^{*} &  c_{-,-1}^{}~c_{-,1}^{*} &  c_{-,-1}^{}~c_{-,-1}^{*} & \\   \label{quin1}
\end{pmatrix}
\equiv 
\begin{pmatrix}
& a_{11} & & a_{12} & a_{13}  &  a_{14} &  \\ \\ 
& a_{21} & & a_{22} & a_{23}  &  a_{24} &  \\  \\
& a_{31} & & a_{32} & a_{33}  &  a_{34} & \\ \\
& a_{41} & & a_{42} & a_{43}  &  a_{44} & \\   \label{quin1}
\end{pmatrix}. 
\end{eqnarray} 
The characteristic equation for the $4\times4$ matrix $\rho^{T_l}$ is given as,

\begin{eqnarray}\nonumber
    && \lambda^4-\lambda^{3}\sum^4_{i=1}|a_{ii}|+ \lambda^2 \sum^{3}_{i=1} \sum^{4}_{j=2}
    \begin{vmatrix}
        a_{ii} && a_{ij}\\
        a_{ji} && a_{jj}
    \end{vmatrix}_{i<j} 
 - \lambda \sum^{2}_{i=1} \sum^{3}_{j=2}\sum^{4}_{k=3} \begin{vmatrix}
        a_{ii} && a_{ij} && a_{ik}\\
        a_{ji} && a_{jj} && a_{jk} \\
        a_{ki} && a_{kj} && a_{kk}
    \end{vmatrix}_{i<j<k}  + |\rho^{T_l}|=0
\end{eqnarray}
 Solving the algebra after substituting the $a_{ij}$ in terms if $c's$, we get the following, 
 \begin{eqnarray}
     \lambda^4-\lambda^3-A\lambda +A^2 = 0, \label{abc}
 \end{eqnarray}
 where $A$ is the determinant of the reduced density matrix:
 \begin{eqnarray}
\begin{vmatrix}
        a_{11} + a_{22}  && a_{13} + a_{24} \\
        a_{31} + a_{42} &&  a_{33} + a_{34}  
    \end{vmatrix}. 
\end{eqnarray} 
 Four roots of the equation give the four eigenvalues of the partially transposed density matrix $\rho^{T_l}$.  \\

{\color{Red} $\bullet$} {\color{NavyBlue} {\it $2\otimes3$ system}:~}
We take the basis for the smaller subsystem as $\{+,-\}$ and other as $\{+1,0,-1\}$. 
The density matrix for the $2\otimes3$ composite system can be written as,
\begin{eqnarray}
\rho = 
\begin{pmatrix}
& c_{+,1}^{}~c_{+,1}^{*} & c_{+,1}^{}~c_{+,0}^{*} & c_{+,1}^{}~c_{+,-1}^{*} & c_{+,1}^{}~c_{-,-1}^{*}  & c_{+,1}^{}~c_{-,0}^{*}  &  c_{+,1}^{}~c_{-,-1}^{*} &  \\ \\ 

& c_{+,0}^{}~c_{+,1}^{*} & c_{+,0}^{}~c_{+,0}^{*} & c_{+,0}^{}~c_{+,-1}^{*} & c_{+,0}^{}~c_{-,-1}^{*}  & c_{+,0}^{}~c_{-,0}^{*}  &  c_{+,0}^{}~c_{-,-1}^{*}  &  \\  \\

& c_{+,-1}^{}~c_{+,1}^{*} & c_{+,-1}^{}~c_{+,0}^{*} & c_{+,-1}^{}~c_{+,-1}^{*}  & c_{+,-1}^{}~c_{-,-1}^{*} &  c_{+,-1}^{}~c_{-,0}^{*} & c_{+,-1}^{}~c_{-,-1}^{*}  &  \\  \\              
& c_{-,1}^{}~c_{+,1}^{*} & c_{-,1}^{}~c_{+,0}^{*} & c_{-,1}^{}~c_{+,-1}^{*}  & c_{-,1}^{}~c_{-,-1}^{*} &  c_{-,1}^{}~c_{-,0}^{*} & c_{-,1}^{}~c_{-,-1}^{*}  & \\ \\

& c_{-,0}^{}~c_{+,1}^{*} & c_{-,0}^{}~c_{+,0}^{*} & c_{-,0}^{}~c_{+,-1}^{*}  & c_{-,0}^{}~c_{-,-1}^{*} &  c_{-,0}^{}~c_{-,0}^{*} & c_{-,0}^{}~c_{-,-1}^{*}  & \\ \\

& c_{-,-1}^{}~c_{+,1}^{*} & c_{-,-1}^{}~c_{+,0}^{*} & c_{-,-1}^{}~c_{+,-1}^{*}  & c_{-,-1}^{}~c_{-,-1}^{*} &  c_{-,-1}^{}~c_{-,0}^{*} & c_{-,-1}^{}~c_{-,-1}^{*}   & 
\label{quin1}
\end{pmatrix}.
\end{eqnarray} 
The partial transpose of the above density matrix will then be,
\begin{eqnarray}
\rho^{T_l} &=&
\begin{pmatrix}
& c_{+,1}^{}~c_{+,1}^{*} & c_{+,0}^{}~c_{+,1}^{*}  & c_{+,-1}^{}~c_{+,1}^{*} & c_{+,1}^{}~c_{-,-1}^{*}  &  c_{+,0}^{}~c_{-,-1}^{*}  & c_{+,-1}^{}~c_{-,-1}^{*}  &  \\ \\ 

& c_{+,1}^{}~c_{+,0}^{*} & c_{+,0}^{}~c_{+,0}^{*} & c_{+,-1}^{}~c_{+,0}^{*} & c_{+,1}^{}~c_{-,0}^{*}  & c_{+,0}^{}~c_{-,0}^{*}  &  c_{+,-1}^{}~c_{-,0}^{*}  &  \\  \\

&  c_{+,1}^{}~c_{+,-1}^{*} &  c_{+,0}^{}~c_{+,-1}^{*}  & c_{+,-1}^{}~c_{+,-1}^{*}  &  c_{+,1}^{}~c_{-,-1}^{*} &   c_{+,0}^{}~c_{-,-1}^{*} & c_{+,-1}^{}~c_{-,-1}^{*}  &  \\  \\              
& c_{-,1}^{}~c_{+,1}^{*} & c_{-,0}^{}~c_{+,1}^{*} & c_{-,-1}^{}~c_{+,1}^{*}   & c_{-,1}^{}~c_{-,-1}^{*} & c_{-,0}^{}~c_{-,-1}^{*}   & c_{-,-1}^{}~c_{-,-1}^{*}  & \\ \\

&  c_{-,1}^{}~c_{+,0}^{*} & c_{-,0}^{}~c_{+,0}^{*} & c_{-,-1}^{}~c_{+,0}^{*}  & c_{-,1}^{}~c_{-,0}^{*} &  c_{-,0}^{}~c_{-,0}^{*} & c_{-,-1}^{}~c_{-,0}^{*}   & \\ \\

& c_{-,1}^{}~c_{+,-1}^{*} &  c_{-,0}^{}~c_{+,-1}^{*} & c_{-,-1}^{}~c_{+,-1}^{*}  &  c_{-,1}^{}~c_{-,-1}^{*} &  c_{-,0}^{}~c_{-,-1}^{*} & c_{-,-1}^{}~c_{-,-1}^{*}   & 
\label{quin1}
\end{pmatrix}
\equiv 
\begin{pmatrix}
& a_{11} &  a_{12} & a_{13}  &  a_{14} & a_{15}  &  a_{16} & \\  \\ 
& a_{21} &  a_{22} & a_{23}  &  a_{24} & a_{25}  &  a_{26} & \\  \\
& a_{31} &  a_{32} & a_{33}  &  a_{34} & a_{35}  &  a_{36} & \\  \\
& a_{41} &  a_{42} & a_{43}  &  a_{44} & a_{45}  &  a_{46} & \\  \\  
& a_{51} &  a_{52} & a_{53}  &  a_{44} & a_{55}  &  a_{56} & \\  \\ 
& a_{61} &  a_{62} & a_{63}  &  a_{64} & a_{65}  &  a_{66} & 
\label{quin1}
\end{pmatrix}. 
\end{eqnarray} 
The characteristic equation for the $6\times6$ matrix $\rho^{T_l}$ is given as,

\begin{eqnarray}\nonumber
    && \lambda^6-\lambda^{5}\sum^6_{i=1}|a_{ii}|+ \lambda^4 \sum^{5}_{i=1} \sum^{6}_{j=2}
    \begin{vmatrix}
        a_{ii} && a_{ij}\\
        a_{ji} && a_{jj}
    \end{vmatrix}_{i<j} 
 - \lambda^3 \sum^{4}_{i=1} \sum^{5}_{j=2}\sum^{6}_{k=3} \begin{vmatrix}
        a_{ii} && a_{ij} && a_{ik}\\
        a_{ji} && a_{jj} && a_{jk} \\
        a_{ki} && a_{kj} && a_{kk}
    \end{vmatrix}_{i<j<k<m} 
     + \lambda^2 \sum^{3}_{i=1} \sum^{4}_{j=2} \sum^{5}_{k=3}  \sum^{6}_{m=4}  \begin{vmatrix}
        a_{ii} && a_{ij} && a_{ik} && a_{im} \\
        a_{ji} && a_{jj} && a_{jk} && a_{im} \\
        a_{ki} && a_{kj} && a_{kk} && a_{im} \\
        a_{ki} && a_{kj} && a_{kk} && a_{im} \\ 
    \end{vmatrix}_{i<j<k<m}    \\ 
    && ~~~~~~~~~~~~~~~~  + \lambda \sum^{2}_{i=1} \sum^{3}_{j=2} \sum^{4}_{k=3}  \sum^{5}_{m=4}  \sum^{6}_{m=5}  \begin{vmatrix}
        a_{ii} && a_{ij} && a_{ik} && a_{im} && a_{in} \\
        a_{ji} && a_{jj} && a_{jk} && a_{im} && a_{in} \\
        a_{ki} && a_{kj} && a_{kk} && a_{im} && a_{in} \\
        a_{ki} && a_{kj} && a_{kk} && a_{im} && a_{in} \\ 
        a_{ki} && a_{kj} && a_{kk} && a_{im} && a_{in} \\ 
    \end{vmatrix}_{i<j<k<m<n} 
     + |\rho^{T_l}|=0
\end{eqnarray}
As the power of $\lambda$ gets down, the dimension of the determinants in their coefficients increases; however, the number of such determinants will decrease, $e.g.$ coefficient of $\lambda^4$ contains $\binom{6}{2}=15$ determinants of $2 \times 2$ matrices, coefficient of $\lambda^3$ contains $\binom{6}{3}=30$ determinants of $3 \times 3$ matrices, coefficient of $\lambda^2$ contains $\binom{6}{4}=15$ determinants of $4 \times 4$ matrices, $\lambda$ contains $\binom{6}{5}=6$ determinants of $5 \times 5$ matrices. All determinants of $5 \times 5$ matrices vanish. Except for a few, most of the other determinants are identically zero. Therefore, the calculation is manageable. 
 Solving the algebra after substituting the $a_{ij}$ in terms of $c_{\pm,l}$, we see that only $\lambda^6$, $\lambda^5$, $\lambda^3$, $\lambda^2$ terms survive, rest are zero, leading to,  
 \begin{eqnarray}
     \lambda^6-\lambda^5-A\lambda^3 +A^2\lambda^2 = 0, \label{abc}
 \end{eqnarray}
This can further be written as, 
 \begin{eqnarray}
     \lambda^2(\lambda^4-\lambda^3-A\lambda +A^2) = 0, \label{abc}
 \end{eqnarray}
which shows four non-zero and two trivial zero eigenvalues. Here, $A$ is the determinant of the reduced density matrix:
 \begin{eqnarray}
\begin{vmatrix}
        a_{11} + a_{22} + a_{33}  && a_{14} + a_{25} + a_{36} \\
        a_{41} + a_{52} + a_{63} &&  a_{63} + a_{55} + a_{66}
    \end{vmatrix}. 
\end{eqnarray} \\

{\color{Red} $\bullet$} {\color{NavyBlue} {\it $2\otimes n$ system}:~}
The characteristic equation for the general $n \times n$ matrix is, 
\begin{eqnarray}\nonumber
    && (-\lambda)^n+ (-\lambda)^{n-1}\sum^n_{i=1}|a_{ii}|+ (-\lambda)^{n-2} \sum^{n-1}_{i=1} \sum^{n}_{j=2}
    \begin{vmatrix}
        a_{ii} && a_{ij}\\
        a_{ji} && a_{jj}
    \end{vmatrix}_{i<j} 
  +(-\lambda)^{n-3} \sum^{n-2}_{i=1} \sum^{n-1}_{j=2}\sum^{n}_{k=3} \begin{vmatrix}
        a_{ii} && a_{ij} && a_{ik}\\
        a_{ji} && a_{jj} && a_{jk} \\
        a_{ki} && a_{kj} && a_{kk}
    \end{vmatrix}_{i<j<k<m}  \nonumber  \\    && (-\lambda)^{n-4} \sum^{n-3}_{i=1} \sum^{n-2}_{j=2} \sum^{n-1}_{k=3}  \sum^{n}_{m=4}  \begin{vmatrix}
        a_{ii} && a_{ij} && a_{ik} && a_{im} \\
        a_{ji} && a_{jj} && a_{jk} && a_{im} \\
        a_{ki} && a_{kj} && a_{kk} && a_{im} \\
        a_{ki} && a_{kj} && a_{kk} && a_{im} \\ 
    \end{vmatrix}_{i<j<k<m}  
      +(-\lambda)^{n-5} \sum^{n-4}_{i=1} \sum^{n-3}_{j=2} \sum^{n-2}_{k=3}  \sum^{n-1}_{m=4}  \sum^{n}_{m=5}  \begin{vmatrix}
        a_{ii} && a_{ij} && a_{ik} && a_{im} && a_{in} \\
        a_{ji} && a_{jj} && a_{jk} && a_{im} && a_{in} \\
        a_{ki} && a_{kj} && a_{kk} && a_{im} && a_{in} \\
        a_{ki} && a_{kj} && a_{kk} && a_{im} && a_{in} \\ 
        a_{ki} && a_{kj} && a_{kk} && a_{im} && a_{in} \\ 
    \end{vmatrix}_{i<j<k<m<n} \nonumber \\
  &&  ~~~~~~~~~~~~~~~~~~~~  +
    ... 
     + |\rho^{T_l}|=0
\end{eqnarray}
For the partially transposed density matrix acting on $2 \otimes n$, all determinants of matrices $5 \times 5$, or higher dimensions, that appear are identically zero. This will lead to the following characteristic equation, 
 \begin{eqnarray}
     \lambda^{2n}-\lambda^{2n-1}-A\lambda^{2n-3} +A^2\lambda^{2n-4} = 0, \label{abc}
 \end{eqnarray}
This can further be written as, 
 \begin{eqnarray}
     \lambda^{2n-4}(\lambda^4-\lambda^3-A\lambda +A^2) = 0.  \label{abc}
 \end{eqnarray}

\end{widetext}

\end{document}